\documentclass[table]{spie}  %>>> use for US letter paper
\newcommand{\tentoe}{\ensuremath{10^{-8}}}

\newcommand{\tentot}{\ensuremath{10^{-10}}}

\newcommand{\vvcP}{\ensuremath{e^{il\theta}}}
\newcommand{\vvcM}{\ensuremath{e^{-il\theta}}}

\usepackage{amsmath,amsfonts,amssymb}
\usepackage{graphicx}
\usepackage[colorlinks=true, allcolors=blue]{hyperref}
\usepackage{orcidlink} % requires hyperref and tikz

\title{New method to achieve the proper polarization state for a vector vortex coronagraph}

\author[a]{Jorge Llop-Sayson\orcidlink{0000-0002-3414-784X}}
\author[b]{Cole~Kappel}
\author[a]{Nemanja~Jovanovic}
\author[a,c]{Dimitri Mawet}
\affil[a]{California Institute of Technology, 1200 E California Blvd., Pasadena, CA 91125}
\affil[b]{University of California, Irvine, Irvine, California 92697}
\affil[c]{Jet Propulsion Laboratory, California Institute of Technology, 4800 Oak Grove Drive, Pasadena, CA 91109}

\authorinfo{Send correspondence to J. Llop-Sayson\\E-mail: {jllopsay@caltech.edu}}

\begin{document} 
\maketitle

\begin{abstract}
The vector vortex coronagraph (VVC) performance in the laboratory and in ground-based observatories has earned it a spot on the NASA mission concepts HabEx and LUVOIR. The VVC induces a phase ramp through the manipulation of the polarization state. Left- and right-circular polarizations get imprinted a phase ramp of opposite signs, which prevents model-based focal plane wavefront sensing and control strategies in natural light. We thus have to work with a polarization state than ensures circularly polarized light at the VVC mask. However, achieving this polarization state can be non trivial if there are optics that add phase retardance of any kind between the circular polarizer and the focal plane mask. Here we present the method currently used at the Caltech high contrast spectroscopy testbed (HCST) to achieve the proper circular polarization state for a VVC, which only uses the deformable mirror and appropriate rotation of the circular polarizer and analyzer optics. At HCST we achieve raw contrast levels of \tentoe~for broadband light with a VVC. 
\end{abstract}

% Include a list of keywords after the abstract 

\section{INTRODUCTION}
\label{sec:intro}  % \label{} allows reference to this section
The vortex coronagraph (VC) concept\cite{Foo2005,Mawet2005} has enabled the imaging of exoplanets from ground based telescopes for their characterization \cite{Wang2020,Wagner2021,llopsayson2021}. Its performance in terms of inner working angle, contrast, throughput to off-axis sources and sensitivity to low order aberrations makes the VC an excellent coronagraph architecture for most cases\cite{Ruane2017,Ruane2018habex}. The NASA mission concepts considered by the Astronomy and Astrophysics Decadal Survey (Astro2020) HabEx\cite{HabExFinalReport} and LUVOIR-B\cite{LUVOIRfinalReport} have the VC as their baseline coronagraph mode.

A VC imprints an azimuthal phase ramp on the beam at a focal plane of the form $e^{\pm i l \theta}$, where $l$ is the charge and $\theta$ the azimuthal angle.
For an on-axis source, i.e. the star, this optical vortex diffracts all the light out of the outer edge of the beam at the following pupil plane, where a Lyot stop blocks the diffracted light. An off-axis source, such as a planet, avoids the singularity at the center of the vortex and is transmitted through the Lyot stop. To fabricate an optical vortex we resort to polarization changes in order to build an achromatic vortex phase ramp. We refer to this kind of vortex as a vector vortex coronagraph (VVC)\cite{Mawet2009vvcDemo}. Some manufacturing techniques that successfully build achromatic VVCs are: liquid crystal polymers\cite{Mawet2009vvcTheory}, subwavelength gratings \cite{Mawet2005_subwavelengthGratings,Catalan2016}, and photonic crystals \cite{Murakami2013}. The VVC geometrical phase has an important limitation: the phase ramp is imprinted with a different sign for left- and right-circular polarizations. We are currently researching ways to manufacture achromatic scalar vortex coronagraphs (see Desai et al. in these proceedings). 

Furthermore, the manufacturing of a VVC is not perfect; a retardance error is introduced that causes a chromatic polarization leakage term, for which the vortex ramp is not applied. Hence, this term goes through the Lyot stop freely and ends in the image plane. All these challenges require our close attention on the polarization state of our system. In this work we are interested in ensuring that we work with the proper polarization state in order to do wavefront control and achieve the best performance a VVC has to offer. 

%%%%%%%%%%%%%%%%%%%%%%%%%%%%%%%%%%%%%%%%%%%%%%%%%%%%%%%%%%%%%%%%%%%%%%%%%%%%%%
\section{Achieving the proper polarization state for a VVC}
\subsection{The vector vortex coronagraph polarization model}
\label{sec:math}
Due to the physical process that takes place at the focal plane mask of a VVC and how the vortex ramp is imprinted with polarization changes, the VVC mask acts as a half-wave plate that imprints a vortex ramp of different sign for the left- and right-circular polarizations. A VVC mask in matrix form using the Jones matrix formalism in circular basis is written as:
\begin{equation}\label{eq:vvc_ideal}
FPM_{VVC, ideal} = 
\begin{bmatrix}
0 & \vvcP \\
\vvcM & 0 
\end{bmatrix}
\end{equation}

Where $\theta$ is the azimuthal angle around the axis. The right-circular polarization going through the VVC mask will become left-circular with a positive vortex ramp; the left-circular will become right-circular with the opposite sign ramp. An idealized VVC, as shown in Eq.~\ref{eq:vvc_ideal}, does not account for limitations associated with manufacturing imperfections of the focal plane mask. In practice, a VVC has a non-ideal transmissivity and non-zero polarization leakage; in matrix form:

\begin{equation}\label{eq:vvc_leak}
FPM_{VVC} = c_V 
\begin{bmatrix}
0 & \vvcP \\
\vvcM & 0 
\end{bmatrix}
+
c_L 
\begin{bmatrix}
1 & 0 \\
0 & 1 
\end{bmatrix},
\end{equation}
where $c_V$ and $c_L$ are constants associated to the transmissivity and polarization leakage respectively. We refer to Ref.~\citenum{ruane2020achromaticvvc} for a comprehensive and thorough  description of the fundamentals of a VVC. 

The leakage term amplitude, $c_L$, is driven by the manufacturing quality of the focal plane mask, but is, however, inevitable. This incoherent term is thus not affected by the Lyot stop since it does not carry the vortex ramp. The only way of cancelling it is by adding a circular analyzer downstream of the focal plane mask, a technique often referred to as polarization filtering. 

\subsection{The proper polarization state}
The fact that the two orthogonal polarizations have different phase ramps (see Eq.~\ref{eq:vvc_ideal}) makes it impossible to distinguish them to do focal plane wavefront control with our current methods. Indeed, to do wavefront control an estimate of the electric field is needed at the image plane, and to get such estimate we apply phase diversity that inevitably gets confused if both polarizations are present at the focal plane mask. There is no way of effectively doing wavefront control if left and right-circular polarizations have comparable intensities at the focal plane mask, these two states would be incoherent in the image plane. Therefore, we need circular polarization at the focal plane mask of a VVC.

Therefore, for an unpolarized source, a VVC requires a circular polarizer (or beamsplitter) that sets the polarization state of the beam to pure circular polarization at the focal plane mask. The leakage term is then dealt with with a circular analyzer downstream of the mask. Any other configuration results in two incoherent components of the light that cannot be disentangled with phase diversity.

%%%%%%%%%%%%%%%%%%%%%%%%%%%%%%%%%%%%%%%%%%%%%%%%%%%%%%%%%%%%%%%%%%%%%%%%%%%%%%
\subsection{Method to achieve the proper polarization state}
\label{sec:method}
Achieving circular polarization at the focal plane mask can be non trivial. The optics in a circular polarizer and analyzer are transmissive optics, which makes them susceptible of introducing unwanted wavefront errors. In a high contrast imaging setup, the wavefront error budget is usually very strict, so we want to avoid having these optics right next to the focal plane mask. 
Such is the case of HCST. As presented in Sec.~\ref{sec:setup}, our circular polarizer is upstream of the entrance pinhole, in which case it does not introduce any wavefront error thanks to the filtering of the pinhole. Furthermore, the circular analyzer is placed downstream of both the Lyot stop and the field stop, which block most of the leftover light outside of the region of interest. In this configuration the wavefront error that the polarizer and analyzer would introduce is largely eliminated. However, this results in having several optics between the circular polarizer and the focal plane mask, and between the focal plane mask and the circular analyzer.

The optics in the path change the polarization state of the light. This is especially true for flat mirrors at wide angles\cite{Mudge2014}, but even parabolic mirrors add polarization aberrations\cite{Breckinridge2015}. Therefore, even if we set the circular polarizer at the entrance of the setup in such a way that the light is in a circular polarization state, i.e. linear polarizer and quarter-wave plate fast axis at 45$^\circ$, by the time the light arrives at the focal plane mask, the polarization state has changed. In this case we would find a mixing of left and right polarizations that induces the incoherent terms discussed in Sec.~\ref{sec:math}. 

\begin{figure}[t!]
   \begin{center}
   \begin{tabular}{c} %% 
   \includegraphics[height=6.5cm]{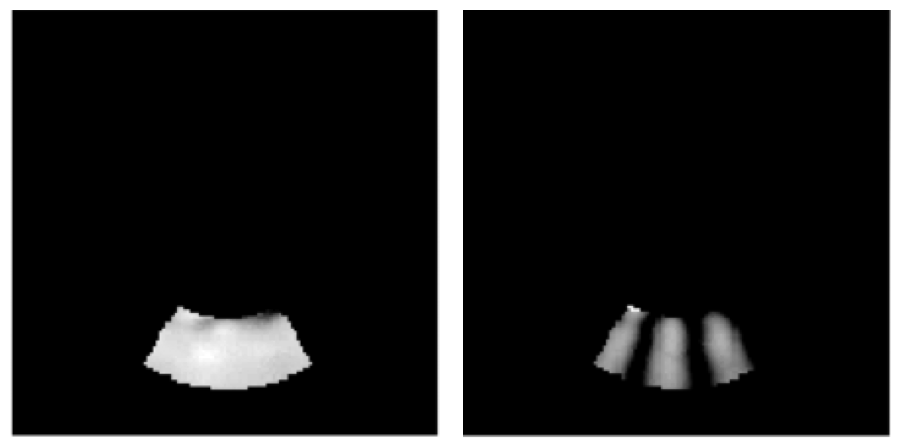}
   \end{tabular}
   \end{center}
   \caption{
   \label{fig:modulation}
   A flat probe is sent with the DM to determine the polarization state at the focal plane mask. Both images show the probe intensity at the image plane. \textit{Left}: the polarization is close to pure circular polarization at the focal plane mask, so, when going through the circular analyzer and arriving on the image plane, the probe remains flat. \textit{Right}: a mixture of left and right polarization is present at the focal plane mask; when going through the circular analyzer an azimuthal modulation becomes apparent in the image plane. } 
\end{figure}

A way to know if the polarization is pure circular at the focal plane mask is by sending a flat probe with a \textit{sinc} function on the deformable mirror (DM). A flat probe remains flat in intensity distribution through the circular analyzer only if there is pure circular polarization at the focal plane mask. If not, the mixture of polarizations going through the circular analyzer optics results in an azimuthal modulation that can be seen in the image plane. In Fig.~\ref{fig:modulation} we show both cases. To show this in theory we resort to a simplified mathematical model of the testbed. We first assume that the light reaches the circular analyzer with both polarizations having the same intensity and equal to one, or, as described by a Jones vector: $in=[\vvcM,\vvcP]'$. If this beam goes through the circular analyzer, the optics of which are set to arbitrary positions, i.e. the linear polarizer and quarter-wave plate fast axis are at arbitrary rotation angles, then the light downstream (for both polarizations) can be written as:

\begin{equation}
out \propto c_1\sin(c_2+l\theta) + i(\cos(c_3-l\theta)+\cos(c_4-l\theta))
\end{equation}

Where $c_{1...4}$ are real-valued constants that are function of the position of the linear polarizer and quarter-wave plate axis. This expression describes an azimuthal modulation of the intensity for the input light, $in$. Changing the relative intensity of the left and right polarization terms only changes the amplitude of the different terms in the expression above, but the same azimuthal modulation is maintained. Such expression holds unless: (1) the vortex mask applies the same phase ramp to all the light, i.e. when there is circular polarization at the focal plane mask, (2) when the linear polarizer and quarter-wave plate fast axis are at 45$^\circ$ at the circular analyzer. When there is circular polarization at the focal plane mask, so all the light gets imprinted the same phase ramp sign, the beam immediately downstream of the circular analyzer is only dependent on $\theta$ by the ramp phase change: $out\propto\vvcM$. We avoid being in case (2) by trying several arbitrary configurations of the circular analyzer optics.

% Figure : modulation 

It is worth noting that we do not have a complete picture of the polarization changes in the testbed; with this method we do not determine the polarization aberrations, the retardance added by each optic, or even the combined retardance upstream and downstream of the focal plane mask. However, these are unimportant for all purposes that are relevant for the case at hand: doing wavefront control. We are thus only interested in having pure circularly polarized light immediately upstream of the focal plane mask.

%%%%%%%%%%%%%%%%%%%%%%%%%%%%%%%%%%%%%%%%%%%%%%%%%%%%%%%%%%%%%%%%%%%%%%%%%%%%%%
\section{Laboratory setup: The high contrast spectroscopy testbed (HCST)}
\label{sec:setup}
The experiments presented here are performed at the high contrast spectroscopy testbed (HCST) at Caltech. HCST saw its first light three years ago \cite{Ruane2018hcst}, and we started by implementing wavefront control with the electric field conjugation (EFC) algorithm \cite{GiveOn2009} reaching levels of \tentoe~raw contrast in broadband light\cite{llopsayson2019spie}. We also implemented a way to improve our Jacobian computation with a System Identification based algorithm\cite{llopsayson2019spie}. We demonstrated the Apodized Vortex Coronagraph concept\cite{llopsayson2020avc} that deals with diffracted light from segmentation in the pupil. Lastly, we presented narrowband results for a modified EFC algorithm for single mode fibers, reaching 1$\times$\tentoe~normalized intensity \cite{llopsayson2020spie}.

\begin{figure}[t!]
   \begin{center}
   \begin{tabular}{c} %% 
   \includegraphics[height=6.5cm]{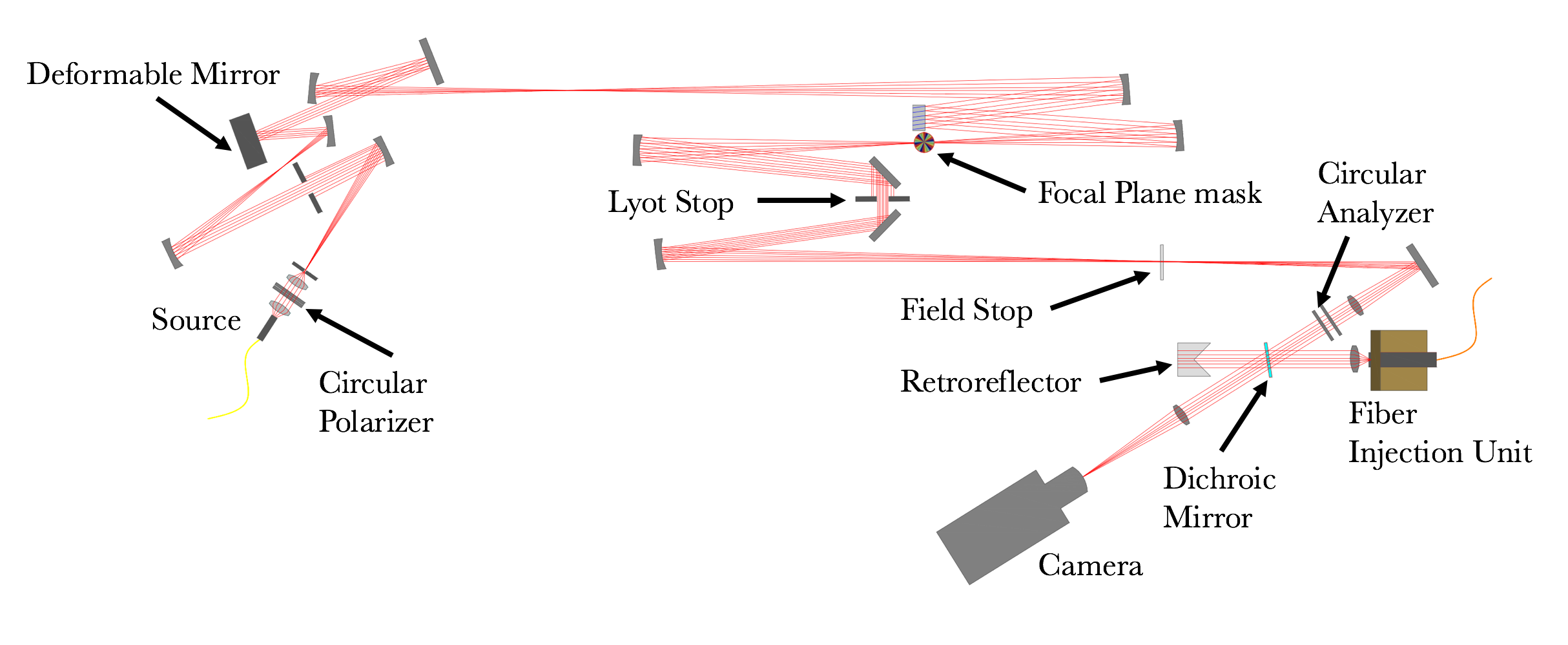}
   \end{tabular}
   \end{center}
   \caption{
   \label{fig:layout}
   HCST's current layout. } 
\end{figure}

In Fig.~\ref{fig:layout} we show the current layout of HCST, and we refer to Ref.~\citenum{llopsayson2020spie} for a complete description of the current laboratory setup. For the work presented here, we note the presence of the circular polarizer and analyzer: the linear polarizers and quarter-wave plates are mounted individually on an actuated rotation stage (CONEX-AG-PR100P) with $\sim$0.001$^\circ$ precision. This allows us to test any combination of configuration in the polarizer/analyzer.
%The linear polarizers are Thorlabs LPVIS100 

% Figure : Laboratory Setup 

%%% Cole
   % - the need: why do we need it
    %- parts: custom arduino board, part number of the sensor, name part of the RTC as well
   % - numbers for sensors: what is the resolution - not accuracy
   
A new feature at HCST is the environmental sensing system. Environmental factors such as temperature, humidity, air turbulence and vibrations affect the data being recorded at HCST. To deal with this, eight sensors have been placed inside the testbed to work with computer programs that display the live environmental data from HCST and any user specified time interval. The environmental sensing system will be used to correlate the data collected with the testbed during any given experiment or time period and the corresponding environmental data. This will help us better understand the limitations of the testbed, and the limitations of current high contrast technologies.
  
At the heart of the environmental sensing system is a custom printed circuit board which attaches to the Arduino hardware and a port that connects to eight sensors. The Arduino hardware board being used is the Adafruit Feather M0 Adalogger which attaches to an Ethernet FeatherWing and a DS3231 Precision Real Time Clock (RTC) FeatherWing. Eight BME680 sensors measure the temperature, humidity and pressure data at the locations that are most sensitive to environmental changes inside the HCST. These sensors measure humidity with 0.008\% r.H. resolution, pressure with 0.18Pa resolution and temperature with 0.01$^\circ$C resolution.

% %%%%%%%%%%%%%%%%%%%%%%%%%%%%%%%%%%%%%%%%%%%%%%%%%%%%%%%%%%%%%%%%%%%%%%%%%%%%%%
% \section{Polarization Crossing at HCST}
% \label{sec:pol}
% \textcolor{red}{yes or no?}

% \begin{figure}[t!]
%   \begin{center}
%   \begin{tabular}{c} %% 
%   \includegraphics[height=5cm]{images/PolarizationCrossing_both.png}
%   \end{tabular}
%   \end{center}
%   \caption{
%   \label{fig:pol}
%   Polarization model of HCST (\texit{left}), measured polarization changes in the testbed (\texit{right}). } 
% \end{figure} 

% \begin{figure}[t!]
%   \begin{center}
%   \begin{tabular}{c} %% 
%   \includegraphics[height=4cm]{images/ProbePolarization.png}
%   \end{tabular}
%   \end{center}
%   \caption{
%   \label{fig:pol}
%   Images from the testbed camera when sending a \textit{sinc} with the DM. The image is centered at the pseudo-star PSF, but we apply a software mask to show just the probed area. The polarization state at the vector vortex coronagraph mask can be deduced to be circular if, when sending a \textit{sinc} probe with the DM, the intensity is not azimuthally modulated. As is the case in the \textit{left} image. On the \textit{right} image, the polarization state can be deduced to be non-circular given the azimuthal modulation.} 
% \end{figure} 

%%%%%%%%%%%%%%%%%%%%%%%%%%%%%%%%%%%%%%%%%%%%%%%%%%%%%%%%%%%%%%%%%%%%%%%%%%%%%%
\section{Laboratory results}
\label{sec:results}
\begin{figure}[t!]
   \begin{center}
   \begin{tabular}{c} %% 
   \includegraphics[height=8cm]{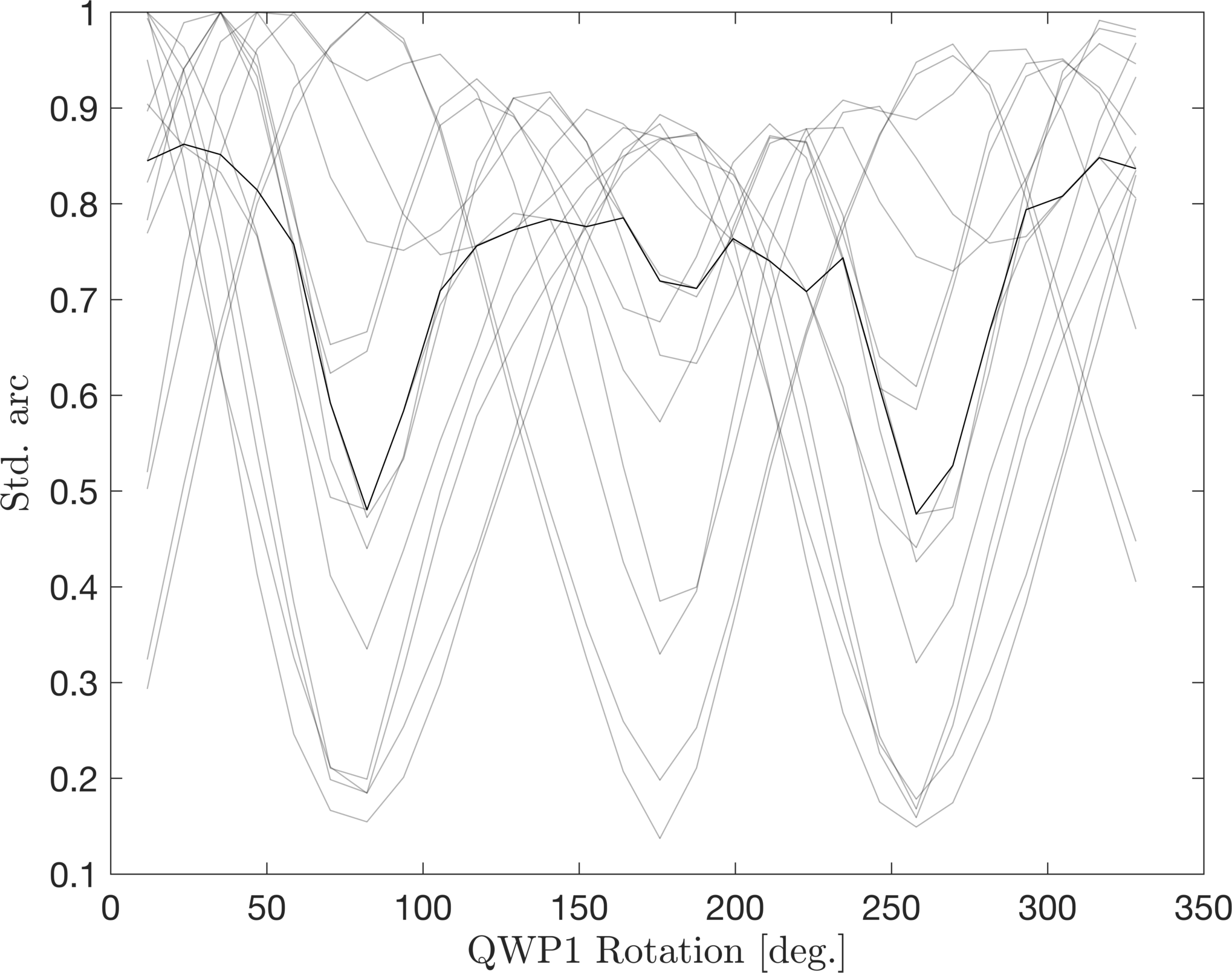}
   \end{tabular}
   \end{center}
   \caption{
   \label{fig:scan_curv}
   The azimuthal standard deviation of the arcs shown in Fig.~\ref{fig:scan_im} indicate the level of azimuthal modulation, and thus how far is the polarization state from pure circular at the focal plane mask. We plot here several curves with different arbitrary configurations for the analyzer, i.e. the positions of its linear polarizer and quarter-wave plate rotation angles, of the modulation intensity versus the rotation of the quarter-wave plate in the circular polarizer at the entrance of the system. \textit{Solid line}: median of all curves. The curves find the same minimum, which corresponds to the position of the polarizer quarter-wave plate that results in circular polarization at the focal plane mask. Four solutions exist (for a 0-360$^\circ$ rotation range).} 
\end{figure}

\begin{figure}[t!]
   \begin{center}
   \begin{tabular}{c} %% 
   \includegraphics[height=21cm]{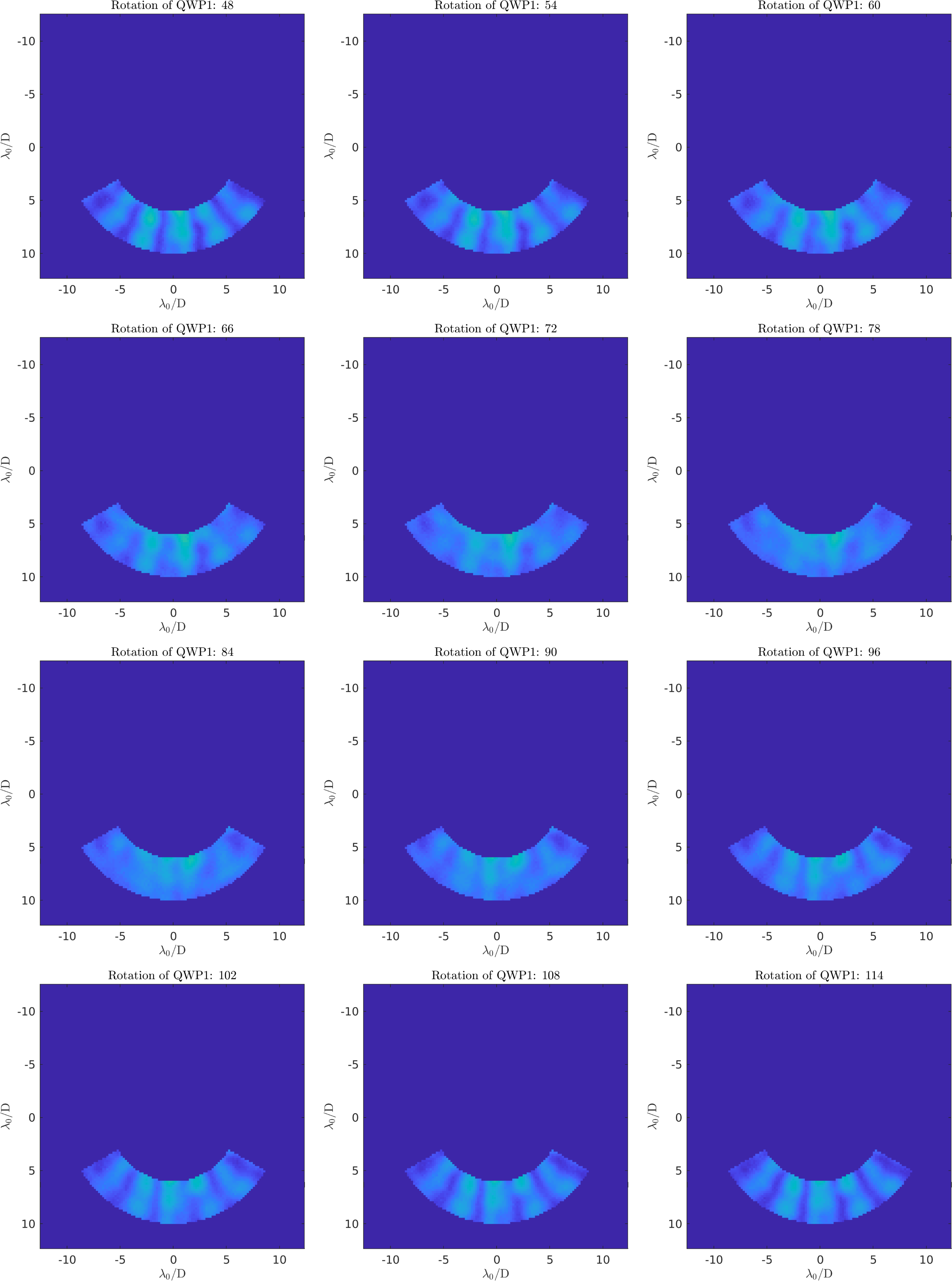}
   \end{tabular}
   \end{center}
   \caption{
   \label{fig:scan_im}
   Images from the testbed camera when sending a \textit{sinc} with the DM. The image is centered at the pseudo-star PSF, but we apply a software mask to show just the probed area. The polarization state at the vector vortex coronagraph mask can be deduced to be circular if, when sending a \textit{sinc} probe with the DM, the intensity is not azimuthally modulated. Different images correspond to a different rotation of the polarizer quarter-wave plate at the entrance, the rotation set is centered at the solution of minimum modulation.} 
\end{figure}

We implement the method described in Sec.~\ref{sec:method} in HCST to achieve the proper polarization state. We rotate the circular polarizer quarter-wave plate at the entrance of the system (see Fig.~\ref{fig:layout}) while we send a flat probe with the DM. We compute the intensity of the probe at the image plane to assess the azimuthal modulation and thus imply how far the polarization state at the focal plane mask is from being circular. To demonstrate this method, we rotate the polarizer quarter-wave plate for different configurations of the circular analyzer at the end of the setup. For each of these configurations we plot the standard deviation of the arc with the probe intensity versus the rotation of the polarizer quarter-wave plate. We show the curves in Fig.~\ref{fig:scan_curv}. In Fig.~\ref{fig:scan_im} we show several images for a set of rotations of the polarizer quarter-wave plate centered at the minimum modulation intensity rotation angle to illustrate how the modulation disappears near the desired polarization state.

% Figure :  

% Figure :  

We pick the polarizer quarter-wave plate that provides no azimuthal modulation, and we compute the contrast between the left and right circular polarizations; we achieve a polarization contrast of a factor of 5$\times$10$^5$ between both polarizations when the circular analyzer is cancelling the leakage term.

%%%%%%%%%%%%%%%%%%%%%%%%%%%%%%%%%%%%%%%%%%%%%%%%%%%%%%%%%%%%%%%%%%%%%%%%%%%%%%
\section{Conclusions}
\label{sec:conclusions}
We present a new method to achieve the proper polarization state for a VVC that does not require the addition of extra optical elements or a polarimeter in the optical path. By sending flat probes with the DM and rotating the optics in the circular polarizer and circular analyzer we can unequivocally determine the polarization state in the system that ensures circular polarization immediately upstream of the focal plane mask. We show how this method works with results from HCST, where we have demonstrated very high contrast levels with the VVC. It is necessary to have the proper polarization state to achieve the raw contrast levels that the VVC promises. The polarization calibration method presented here is particularly relevant to future instruments that envision using the VVC at raw contrast levels of \tentot, such as the HabEx and LUVOIR-B mission concepts.

\acknowledgments % equivalent to \section*{ACKNOWLEDGMENTS}       
The material is based upon work supported by NASA SAT under award No. 80NSSC20K0624.
C. K. is supported by the Caltech WAVE program. J.L.-S. thanks Garreth Ruane for helpful discussions.

% References
\bibliography{report} % bibliography data in report.bib
\bibliographystyle{spiebib} % makes bibtex use spiebib.bst

\end{document}